\documentclass[sigconf,anonymous=false]{acmart}

    \AtBeginDocument{%
  \providecommand\BibTeX{{%
    \normalfont B\kern-0.5em{\scshape i\kern-0.25em b}\kern-0.8em\TeX}}}

\setcopyright{acmcopyright}
\copyrightyear{2019}
\acmYear{2019}

\acmConference[Paris '19]{Paris '19: }{July 25, 2019}{Paris, France}
\acmBooktitle{Paris '19: International Workshop on Explainable Recommendation and Search at ACM-SIGIR}

\usepackage{graphics}
\usepackage[colorinlistoftodos,prependcaption,textsize=tiny]{todonotes}
\usepackage{booktabs}

\setcopyright{rightsretained} 
\acmConference[EARS'19]{SIGIR 2019 Workshop on ExplainAble Recommendation and Search}{July 25, 2019}{Paris, France}

\begin{document}

\title{Model Explanations under Calibration}

\author{Rishabh Jain}
\email{jain.rishabh8897@gmail.com}
\affiliation{%
  \institution{Imperial College London}
}

\author{Pranava Madhyastha}
\email{pranava@imperial.ac.uk}
\affiliation{%
  \institution{Imperial College London}
}

\begin{abstract}
Explaining and interpreting the decisions of recommender systems are becoming extremely relevant both, for improving predictive performance, and providing valid explanations to users. While most of the recent interest has focused on providing local explanations, there has been a much lower emphasis on studying the effects of model dynamics and its impact on explanation. In this paper, we perform a focused study on the impact of model interpretability in the context of calibration. Specifically, we address the challenges of both over-confident and under-confident predictions with interpretability using attention distribution. Our results indicate that the means of using attention distributions for interpretability are highly unstable for un-calibrated models. Our empirical analysis on the stability of attention distribution raises questions on the utility of attention for explainability.
\end{abstract}

\ccsdesc[500]{Information systems~Recommender systems}
\keywords{calibration, explanations, attention, recommender systems, deep learning}


\maketitle

\section{Introduction}

Recommendation systems are used for item filtering based on user preferences in a variety of areas including movies, news, books, social recommendations and products in general. Some commonly used approaches to recommendation systems include  Collaborative filtering, Content-based filtering and hybrid systems. There has been an increased interest in the community in utilizing deep learning based models for recommender systems~\cite{zhang2019deep}. These models can alleviate several limitations of traditional models including complex non-linear transformations, interactions with different types and modalities of inputs. 
Deep learning based models have been particularly shown to be flexible and are known to incorporate additional data when training and can learn from large amounts of auxiliary information, which is usually available to recommendation systems. As deep models are modular than other rigid algorithms they are easily adaptable and extendable. 

The significance of explaining automated recommendations is widely acknowledged~\cite{herlocker2000explaining,tintarev2007survey}. Explanations build user trust, improve their experience, and also give them the opportunity to fix incorrect representations or recommendations. For these reasons, there has been extensive research on ways to explain different types of recommendation systems. We refer the reader to~\citet{zhang2018explainable} for a detailed survey on explainable systems. 

Deep learning based recommendation systems have opened up one way of explaining neural models' outputs in the context of recommendations~\cite{zhang2018explainable} --- by using attention distributions. 
In this context, neural attention mechanisms have gained significant focus, as they have been shown to not only help the model perform better, but also provide explanations by highlighting the input features that play a significant part in computing the model's output~\cite{xue2018deep,Goossens:1999:LWC:553897}. However, it is has been recently indicated that attention may not always provide a reliable form of explanation, especially in the domain of natural language processing~\cite{jain2019analysis}.

One of the emerging problems with the modern neural network models (especially deep neural networks) is their poor calibration~\cite{guo2017calibration}. Over-confident or under-confident predictions can make a model unreliable, especially in sensitive scenarios like health care (disease detection), autonomous driving among others~\cite{guo2017calibration}.

In this paper, we focus on a form of recommendation system that aims to answer \emph{why} a certain recommendation has been made. Especially, we investigate the reliability of attention distributions in deep neural attention based recommendation systems. 

\section{Background and Tools}
In this paper, we investigate the utility of attention with a state-of-the-art deep neural network based model with attention~\cite{xue2018deep}. In this section, we succinctly describe the necessary background and the tools under consideration. 
\subsection{Attention Distribution}
\begin{figure}[h]
  \centering
  \resizebox{.5\linewidth}{!}{ 
  \includegraphics[width=\linewidth,keepaspectratio]{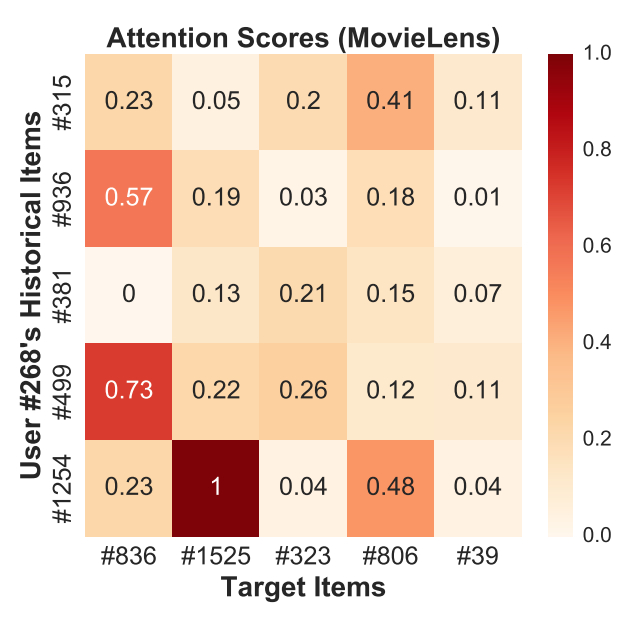}
  }
  \caption{Attention map showing weights assigned to input features}
  \label{fig:atn-map}
  \Description{DeepICF with attention model image.}
\end{figure}

Attention mechanisms, in neural networks, are known to provide the functionality for the model to focus on certain parts of the inputs or features. An attention mechanism in recommendation systems is usually over $u$, a user representation, with the set of item specific representations $\{v_i\} \in \mathcal{V}$ where $\mathcal{V}$ is the domain of all item representations. A compatibility function maps $u$ and $\{v_i\}$ to a scalar distribution, which is then typically converted into a probability distribution using a softmax operator. This usually results in a distribution where some items get more probability mass than others, indicating their influence in the decision made by the system. 
In this paper, we focus on such attention distributions and are interested in their reliability. We are especially interested in understanding the behaviour of the models when the models are mis-calibrated. 
\paragraph{Explanation using Attention}
In neural recommendation systems (and neural networks in general), attention is increasingly being used, not just to improve the model's performance but also as a means to explain the model's predictions~\cite{xue2018deep,wang2018tem,gilpin2018explaining}. The attention maps (heat-maps) are used to indicate which input features to the model were majorly responsible for the model's predictions. In Figure~\ref{fig:atn-map} (from a movie recommendation system from \citet{xue2018deep}), for instance, for target item \#1525, the attention-network assigns the maximum weight to the item \#1254 (one of the previously interacted items of the target user). This information can be used to generate a human-readable explanation like "You are recommended to watch \#1525 because you watched \#1254".

More recently, there has been research on the reliability of attention-maps based explanations~\cite{DBLP:journals/corr/abs-1902-10186} and if they can be used to explain a model. In this paper, we work on this line of research in the context of recommendation systems and their calibration(\ref{calib-section}).
\subsection{Model Calibration}\label{calib-section}
Classification models used as part of any decision process need to be both accurate in their predictions, and should also indicate when they are probably incorrect. 
Model calibration is the degree to which a model's predicted probability correlates with its true correctness likelihood. Calibration measures this property of a model. For example, if a perfectly calibrated model gives 100 different predictions, each with 80\% confidence (probability), 80 of the predictions should be classified correctly.

We use the concept of calibration to plot reliability diagrams~\cite{hamill1997reliability}. A reliability diagram can be defined as the accuracy of the model as a function of its confidence.
%
%
Reliability diagrams help us visualize a model's calibration. A reliability plot which falls below the identity function suggests that the model is over-confident of its predictions (blue plot in Figure \ref{fig:calib-plot}) since it means that the ground truth likelihood (accuracy) is less than the model's confidence in its predictions. On the other hand, it is considered under-confident if the reliability plot is above the identity function. For a perfectly calibrated model, the reliability plot is the identity function.

\subsection{Attention Permutation}\label{atn-perm-sec}
One of the experiments we perform to check the reliability of attention based explanations is permuting the weights randomly and recording the effects of the permutations on the output of the model (inspired from \citet{DBLP:journals/corr/abs-1902-10186}).

Since the particular weights assigned to the input features are used as the basis for the explanations, permuting these weights randomly should cause the model's prediction to change by a substantial margin. In case the predictions remain unchanged it indicates that the attention necessarily doesn't contribute to the predictions. This can be concerning especially when using attention as grounds for explanations.

\subsection{Model Stability}
In our study, we refer to model stability as the consistency of model predictions and internal parameters with different runs of the model by only changing random seeds.
\cite{seed2003uncertainty, seed2007uncertainty}. The seed values are responsible for regulating the training dynamics (weight initialization, training batch generation, among others). This way, we get to measure the impact of these random processes on the output of the model (and the attention weights).


%

\section{Experimental Setup}\label{setup}

In the following sections we describe our experiments and observations. 
\subsection{DeepICF with attention}\label{deepicf}
The DeepICF model uses a deep neural network to learn latent low-dimensional embeddings of users and items that capture implicit and explicit user interactions. 
%
It uses a pair-wise interaction layer, 
which consists of an element-wise product (also called the Hadamard product\cite{wiki:hadamard}) of the target item's latent vector with each of the historical items' vectors. 
The model then follows this with
the pooling layer whose output is a vector of fixed size, to facilitate the deep interaction of layers. This is done via \emph{attention based pooling}. The output of the pooling layer is a vector which condenses the second-order interaction between historical items and the target item (we refer the reader to \citet{xue2018deep} for a detailed explanation of the model).
Finally, the higher order interactions are captured with a multi-layer perceptron. The output of the model is a sigmoid on the final layer's weighted sum.

\emph{Modifications:}
We replaced sigmoid function with a softmax with two outputs over the two classes and trained the model with cross-entropy loss.

\begin{figure}[h]
  \centering
  \resizebox{.9\linewidth}{!}{
  \includegraphics[width=0.9\linewidth,keepaspectratio]{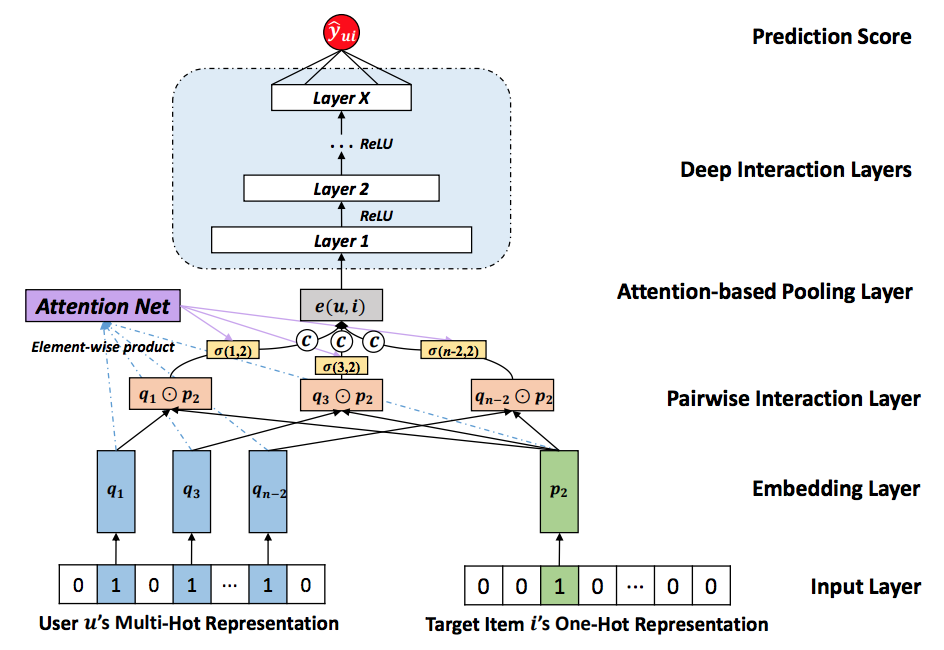}
  }
  \caption{Deep ICF with Attention (from~\citet{xue2018deep})}
  \label{fig:deepicf}
  \Description{DeepICF with attention model image.}
\end{figure}

\subsection{Dataset, Evaluation and Hyperparameters}\label{dataset}
We train, evaluate the model and perform our experiments on the MovieLens\footnote{\url{https://grouplens.org/datasets/movielens/1m/}} dataset. This dataset has been commonly used to evaluate collaborative filtering algorithms. The dataset contains one million ratings where each user has at least 20 ratings and use the standard splits. In our study, we retain the standard procedure used in DeepICF where the original dataset is transformed such that each user item entry is marked as 1: when there is some interaction between the user and item and -1: when there is no interaction the between the target user and item.

\emph{Evaluation:}
For evaluation purposes, the standard metrics used are HitRatio (HR@10) and the NDCG@10(Normalized Discounted Cumulative Gain~\cite{he2015trirank}) as the main metrics. We further use the binary labelling accuracy to investigate the model performance per class, where the classes are defined as: $-1$ when there is no interaction between the user and items and $1$ when there is an explicit interaction (user ratings for the item).
%

\emph{Hyperparameter Settings:}
For training purposes, we use the same hyper-parameters as mentioned in the paper \cite{xue2018deep}. We use the original DeepICF implementation\footnote{\url{https://github.com/linzh92/DeepICF}}.


\section{Results}
Table~\ref{fig:stab-score} compares the performance of the softmax output model with the original DeepICF model and the state-of-the-art Neural Matrix Factorization model\cite{he2017neural}. We observe that the performance of our model is highly competitive and performs as well as the DeepICF with pretraining. In the following sections, we will investigate the reliability of models and the attention distribution in the models. 

\subsection{Calibration}\label{calib}
We plot the reliability diagram for the DeepICF model by bucketing the model predictions based on their confidence and calculating the accuracy for each of the buckets.

\begin{figure}[h]
  \centering
  \resizebox{.9\linewidth}{!}{
  \includegraphics[width=\linewidth,keepaspectratio]{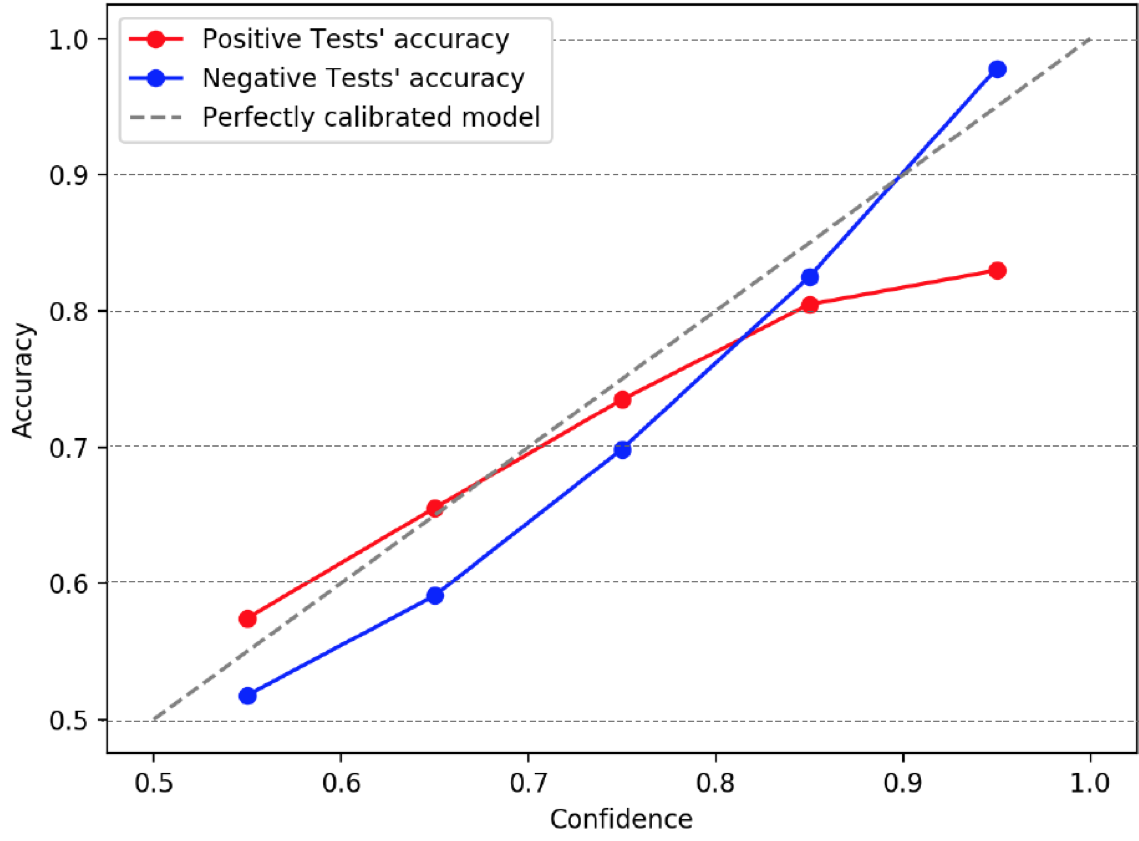}
  }
  \caption{Calibration plot (Reliability diagram) of the DeepICF model.}
  \label{fig:calib-plot}
\end{figure}

We see in Figure~\ref{fig:calib-plot}, for positive test cases, the DeepICF (with attention) model seemingly tends to be over-confident as the confidence increases, where the model tends to be extremely confident about predicting the positive class without being as accurate. This can be problematic especially when dealing with real-world production systems. We also notice that the model is seemingly over-confident in predicting the negative class. This could be because of the imbalance in the dataset where the dataset is extremely skewed towards the negative class. 
We also note that the test-set has a very high degree of imbalance in the number of positive and negative test cases in our test set (1 positive sample for every 99 negative tests). This is impacted in Figure~\ref{fig:calib-plot} as it shows the curves for positive and negative test samples separately. 



\subsection{Attention Permutation}\label{atn_perm}
What is the effect of over-confidence over attention? In order to test the reliability of explanations generated from attention, we permute the attention weights randomly and notice the effect of the permutation on the output of the model (as described in Section~\ref{atn-perm-sec}).

Specifically, in DeepICF, as shown in Figure \ref{fig:deepicf}, the attention based pooling layer assigns a weight for each of the user and item interaction, where the magnitude of the weights indicate the importance of the interaction. In this experiment, we randomly \emph{shuffle} these weights amongst the items and record the difference in the output prediction score (originally classified interaction label). We randomly shuffle the weights $100$ times (as performed in~\citet{DBLP:journals/corr/abs-1902-10186} for each test case, and average the absolute variations in the output predictions.

\begin{figure}[h]
  \centering
  \resizebox{.9\linewidth}{!}{
  \includegraphics[width=\linewidth,keepaspectratio]{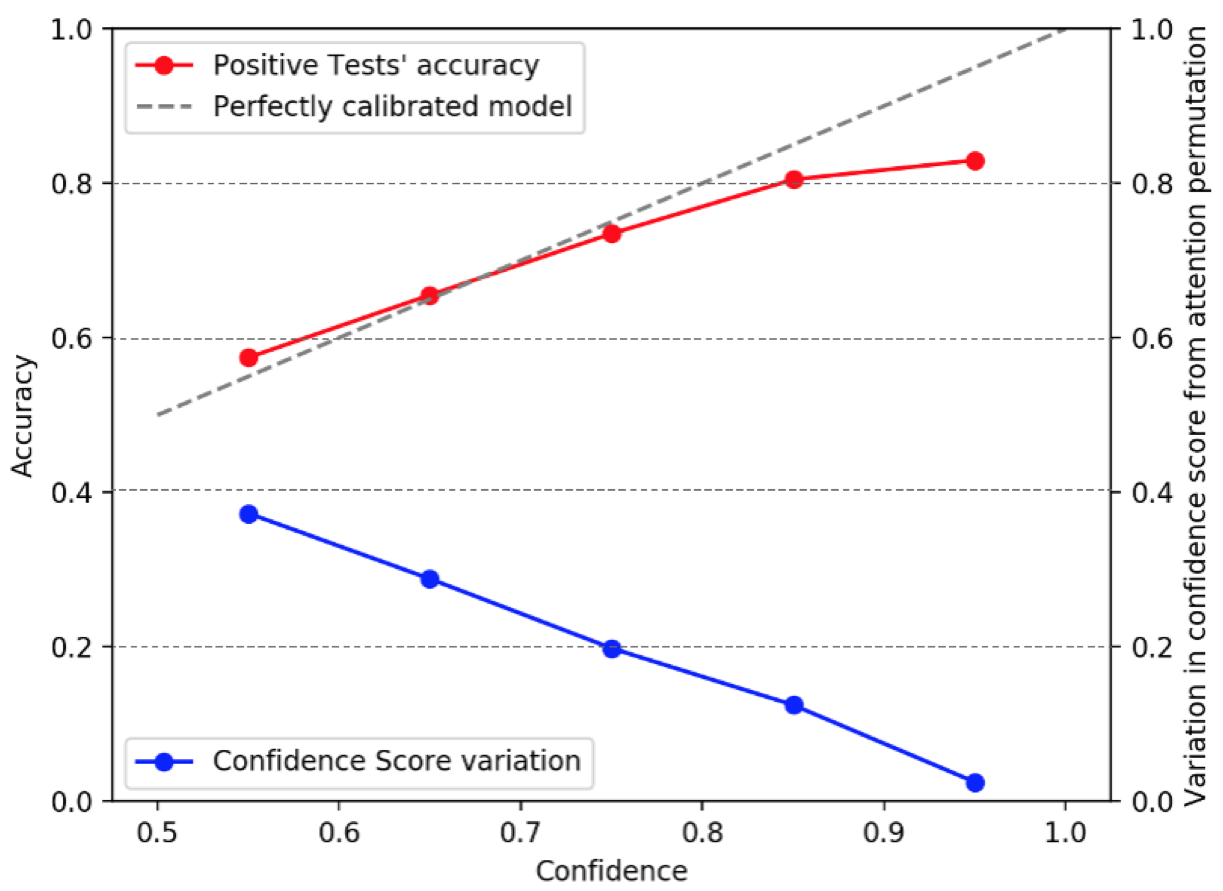}
  }
  \caption{Figure showing the effect of attention permutation (right axis) on the prediction score of wrongly (negatively) classified positive test cases (false negatives).}
  \label{fig:calib-v-atn}
\end{figure}

We plot the average variations in false negatives (right axis) against the confidence of the predicted output for the positive test cases in Figure \ref{fig:calib-v-atn}. We focus on positive test cases as it is the most salient label to measure the model. The plot also contains the reliability diagram for the model (left axis). We note that the perturbations especially have barely any effect on the mis-calibrated cases. In both false positives and false negatives (these increase with mis-calibration), we notice similar trends where the effect of permuting the attention weights decreases as the confidence in the predicted label increases. \textbf{Thus, showing that model explanations generated from the attention distribution become less reliable with over-confident predictions.}

\subsection{Fixing the effect of Class-imbalance}
\begin{figure}[h]
  \centering
  \resizebox{.9\linewidth}{!}{
  \includegraphics[width=\linewidth,keepaspectratio]{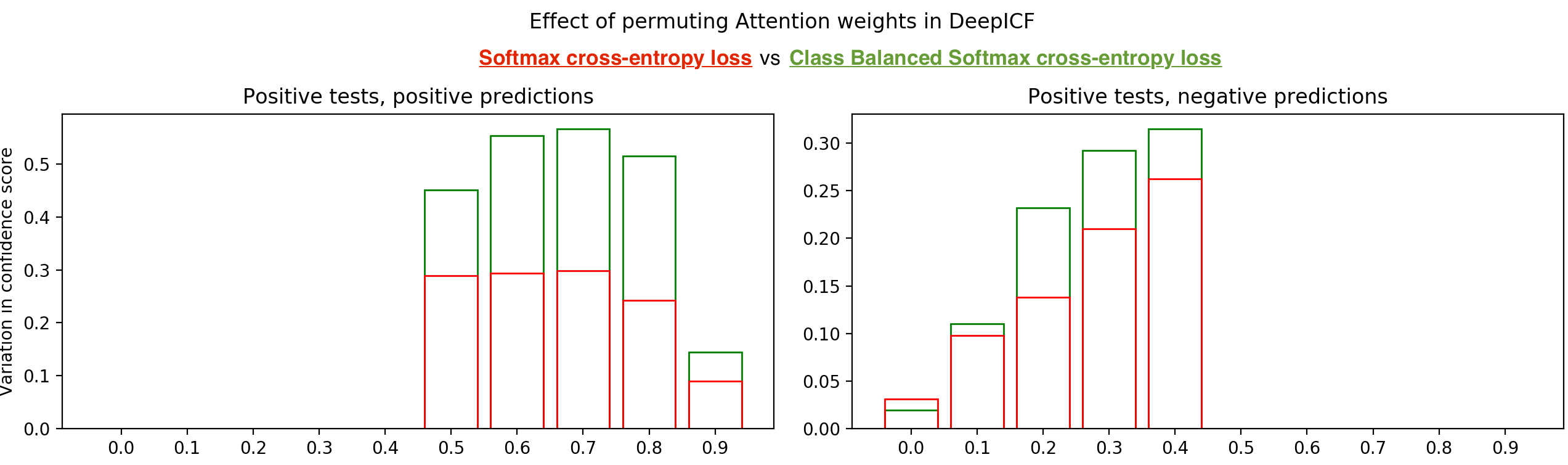}
  }
  \caption{Effect of permuting attention weights.}
  \label{fig:imb}
\end{figure}

As the training split of the dataset is heavily imbalanced: 4 negative labels (no interactions) for every positive label, we use a simple class-weighting heuristic, to cope with this imbalance in the training set and modify our cross-entropy loss. The new loss is calculated by assigning weights to the losses from the test cases such that the loss contribution from both the classes (positive and negative interactions) is balanced\cite{king2001logistic}. We retrain the model with the new loss function and were able to achieve similar HitRatio values to the original model as shown in Table~\ref{fig:stab-score}. We analysed the effect of attention permutation (Section~\ref{atn_perm}) on this model. Figure \ref{fig:imb} compares the new model to the previous model's results. We notice that the new model is considerably more sensitive to attention permutation, compared to the original one. \textbf{This suggests that attention based explanations from the class-balanced loss model are more reliable than the original model.}

\subsection{Stability of DeepICF}
We now consider the effect of random seeds and on initialization of model parameters and in general the model performance. We notice in Table~\ref{fig:stab-score} --- the standard deviation is generally very low suggesting that the performance of the model is seemingly stable and it seems to have small deviation. 
\begin{table}[h]
  \centering
  \begin{tabular}{cll}\toprule
  \textbf{Model type} & \textbf{Hit Ratio@10 (\%)} & \textbf{NDCG@10 (\%)}\\
  \midrule
  DeepICF$^*$ & 68.81 & 41.13 \\
  DeepICF$^*+$Pretrain & 70.84 & 43.80 \\
  NeuMF$^*+$Pretrain & 70.70 & 42.60 \\\midrule
  DeepICF (ours) & 70.41($\pm$0.24) & 43.00 ($\pm$0.34)\\
  DeepICF$+$cls-wt & 68.61 & 41.14
  \\\bottomrule
  \end{tabular}
  \caption{Performance Comparison for DeepICF and NeuMF\cite{he2017neural}. $^*$ indicates scores directly from the corresponding papers. The standard deviation ($\pm$) is obtained with $10$ runs of the model with different random seeds.}
  \label{fig:stab-score}
\end{table}
%
%

\begin{figure}[h]
  \centering
  \resizebox{.9\linewidth}{!}{
  \includegraphics[width=\linewidth,keepaspectratio]{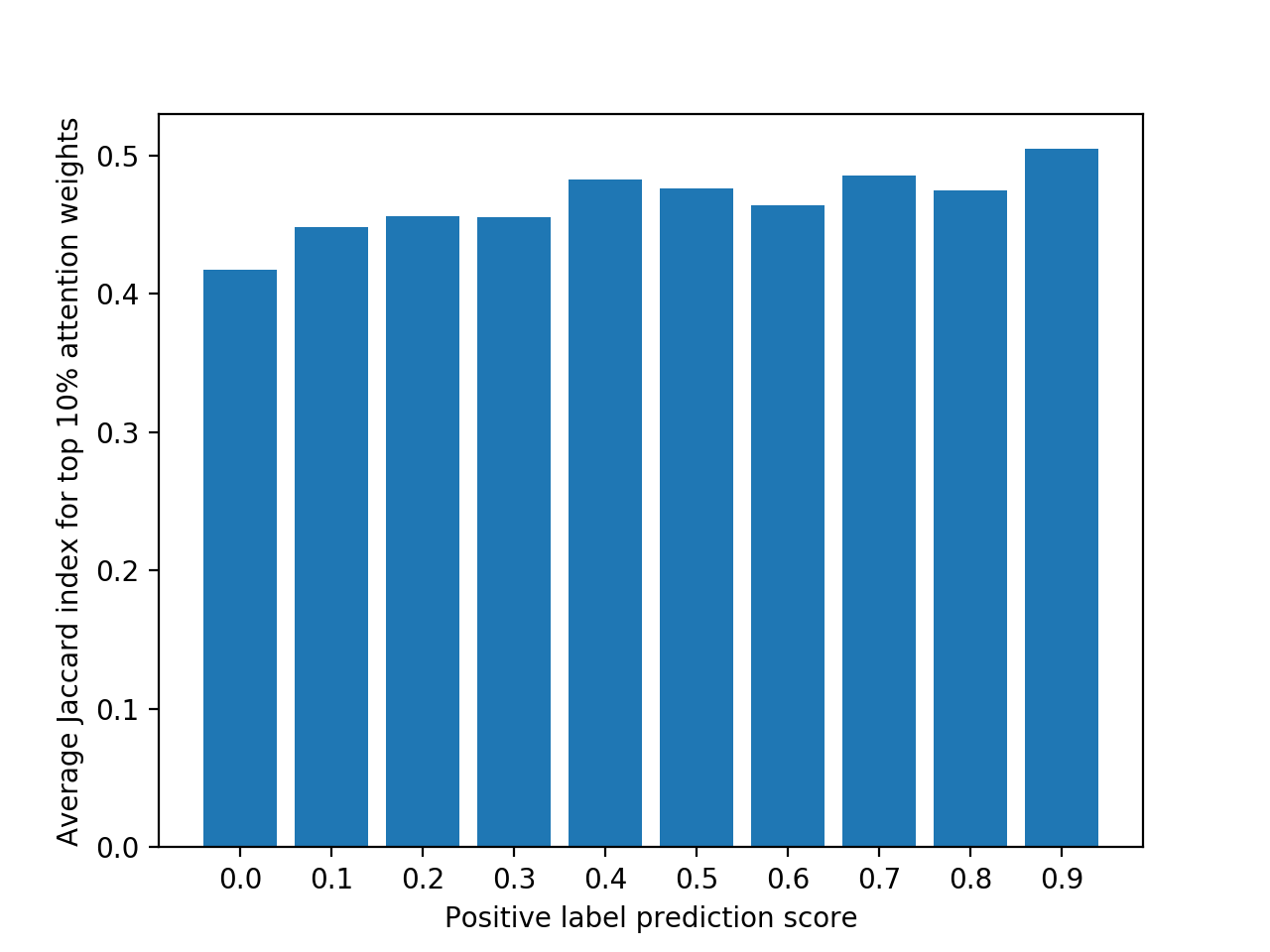}
  }
  \caption{Attention score stability.}
  \label{fig:stable-jaccard}
\end{figure}

\emph{Attention score stability:}
What is the effect of random seeds on attention distribution?  As we are interested in the reliability of attention \emph{explanations}, we focus on the stability of attention scores in DeepICF. We perform the same experiment by running the same model but with $10$ different random seeds and record the \emph{top 10\%} of the most attentive items (user-item interactions which get the highest attention weight assigned) for every particular test case for each model. Then we compare if these top 10 percent most attentive items for a particular test case are consistent for different runs of the models with different random seedst. We calculate the similarity between two sets of items by computing the \texttt{Jaccard Index}~\cite{wiki:jaccard} of the sets. We calculate the Jaccard Index for every possible pair of sets of top attentive items and average over them. Figure \ref{fig:stable-jaccard} shows that the average \texttt{Jaccard Index} for positive predictions with high confidence is around 0.5 (where max \texttt{Jaccard Index} is 1, implying completely stable attention scores).
\textbf{This highlights that the attention explanations from two identical models, trained with different seeds for the same input can vary, severely highlighting the unreliability of such explanations.}

\section{Conclusion}
In this paper, we have explored the importance of model dynamics and its relation to explanation using attention. Concretely, we observe that attention may not be reliable when the selected model is especially mis-calibrated. 
We have explored one possible way of stabilizing the model by accounting for the class imbalance. Significantly, we noticed that using an inverse-class weighted cross-entropy formulation can help improve the stability of attention distribution. 
Further, we observe that over different runs of models with different random seeds, the models seem to obtain different attention distributions. We posit that our work is extremely relevant to the community and can orient towards an important discussion on the reliability of using attention as an explanation. 
%
%
\bibliographystyle{ACM-Reference-Format}
\bibliography{sample-base}
\newpage
\appendix
\vfill
\section*{Appendix}
\textbf{Hyperparameter Settings:}
For training purposes, we use the same model settings for DeepICF as mentioned in the paper \cite{xue2018deep} (or the Github implementation\footnote{\url{https://github.com/linzh92/DeepICF}}), our port of the code is made available at: \url{https://github.com/wakeuprj/DeepICF}.
Hyper-parameters for replication studies are: 
\begin{itemize}
    \item Embedding size: 16
    \item Multi-Layer-Perceptron layers: 64, 32, 16
    \item Alpha ($\alpha$): 0
    \item Beta ($\beta$): 0.8
    \item Learning Rate: 0.01
    \item Pretrain: False
\end{itemize}

\end{document}